\documentclass[12pt,preprint]{aastex}
\def\b{\begin{equation}}
\def\e{\end{equation}}

\shorttitle{Influence of Rotation on Pulsar Radiation}
\shortauthors{Peyman Ahmadi and R. T. Gangadhara}
\received{}

\begin{document}
\title{Influence of Rotation on Pulsar Radiation Characteristics }

\author{ Peyman Ahmadi \altaffilmark{1} and R. T. Gangadhara \altaffilmark{2}}

\affil{Indian Institute of Astrophysics, Bangalore-560034, India}

\altaffiltext{1}{E-mail: {\tt peyman@iiap.ernet.in }}
\altaffiltext{2}{E-mail: {\tt ganga@iiap.ernet.in}}

\begin{abstract}

 	We present a relativistic model for pulsar radio emission  
by including the effect of rotation on coherent curvature 
radiation by bunches. We find that rotation broadens the width of leading
component compared to the width of trailing component. We estimate the 
component 
widths in the average pulse profiles of about 24 pulsars, and find that 
19 of them have a broader leading component. We explain this difference 
in the component widths by using the nested cone emission geometry.

We estimate the effect of pulsar spin on the Stokes 
parameters, and find that the inclination between the rotation and 
magnetic axes can introduce an asymmetry in the circular polarization
of the conal components. We analyze the single pulse polarization
data of PSR~B0329+54 at 606 MHz, and find that in its conal components,
one sense of circular polarization dominates in the leading component
while the other sense dominates in the trailing component.
Our simulation shows that changing the sign of the impact 
parameter changes the sense of circular polarization as well 
as the swing of polarization angle.
\end{abstract}
\keywords{ Radiation mechanism; Rotation; Stokes parameters; Pulsars}

\section{Introduction}

Although a lot of effort has been devoted for understanding the pulsar 
radiation mechanism, it seems, we still do not have a unified model 
for the emission mechanism and the beam structure. Three major  
models have been proposed to explain the
coherent radio emission from pulsars:
emission by bunches (e.g., Komesaroff 1970; Sturrock 1971; Tademaru 1971;
Ruderman and Sutherland 1975; Buschauer and Benford 1976, 1980;
Michel 1978), relativistic plasma emission (e.g., Melrose and Gedalin 1999; 
Asseo and Riazuelo 2000) and maser mechanisms (e.g., Yihan et~al. 1994; Malov 
and Chugunov 1995). Most of them make use of a secondary pair plasma and place 
the origin of the pulsar radiation at the inner region of the magnetosphere. 
 However, the polarization observations (e.g., Clark and Smith 1969; 
Blaskiewicz et~al. 1991) tend to favor the curvature radiation 
as the emission mechanism.  But, the coherent curvature emission by
bunches has been criticized (e.g., Melrose 1981) for its failure to explain 
the creation and stability of bunches.

The rotation vector model (RVM)
by Radhakrishnan and Cook (1969) has been used to interpret the
average polarization angle swing; it assumes a strong dipole magnetic field 
and collimated relativistic flow of plasma. The emitted radiation is then
polarized along or orthogonal to the curvature of the magnetic
field and significantly polarized. 

Manchester et~al. (1975) and Stinebring et~al. (1984 a\&b) have
discussed polarization characteristics of single pulses and
found them highly polarized, with generally the linear
polarization dominating over the circular polarization. They noted that the
circular polarization sense reversal appears to occur close to
the center of the sub-pulse. This signature is not frequency
dependent, contrary to the predictions of propagation or
plasma emission processes (Gangadhara et~al. 1999; Gangadhara and Krishan
1993, 1995).  Rankin (1983) has studied Stokes
parameters for a sample of pulsars, and showed that the
circular polarization patterns are not symmetric, specially for 
core dominant pulsars.
Further, Radhakrishnan and Rankin (1990), in their
phenomenological study of the polarization properties of
pulsars, clarified two extreme types of circular polarization
signature in the average profiles: (a) an anti--symmetric type
wherein the circular polarization changes sense in mid pulse,
and (b) a symmetric type wherein it is predominantly of one
sense. They found a strong correlation between the sense of
position angle swing and change in circular polarization
sense. 

Rankin (1983) suggested that core and conal emissions have
different emission mechanisms, with circular polarization being
a property of core emission only. Lyne and Manchester~(1988)
suggested that a gradual change in emission characteristics from
the core region to the outer edge of the emission beam can
accommodate observations better than two distinct emission
processes. Han et~al. (1998) have studied the circular
polarization in pulsar integrated profiles, and found that
the circular polarization is stronger in the central or 'core'
regions of the pulses, but it is not confined only to this region
as Rankin (1983) claimed.  

Although the theoretical understanding of the polarization properties
of pulsar radiation is a fascinating subject, the
complexity of the details has eluded a major breakthrough so far. 
It seems impossible to accommodate all the
diverse properties of the polarization within a single radiation
mechanism. Gil and Snakowski (1990) have attempted
to examine the polarization properties of the curvature radiation
without taking into account the role of rotation. Gangadhara
(1996) has estimated the energy of particles moving along
rotating magnetic fields but has not estimated the polarization of
the emitted radiation.
Blaskiewicz et~al. (1991) have developed a model by taking into
account the effect of rotation on the particle motion. They have
assumed a constant emission height and estimated the effect of rotation 
only on the position angle swing. 

In this paper, we further develop the mechanism of the curvature emission by
including the rotation and coherency effects, and
estimating the polarization parameters of the emitted radiation. 
In this treatment, we relax the assumption of
a fixed emission height made by Blaskiewicz et~al. (1991). We
show, instead that the major observed features of the circular polarization 
can be explained by considering the emission from extended region. In Sec.~2
 we introduce the equation of motion for individual particles and 
derive an expression for the radiation electric field. In Sec.~3 we 
estimate the Stokes parameters for coherent radiation 
electric field from bunches of plasma particles, by
assuming that the bunches are in an instantaneous circular motion along 
the rotating magnetic field. In Sec.~4 we give the observational 
evidences in favor of predictions of our model.
  
\section{Coherent Radiation from plasma}

The curvature radiation from high energy plasma particles in a strong
curved magnetic field is often postulated to be the mechanism for producing
radiation in the pulsar magnetosphere. The equation for particle dynamics 
is given by
\begin{equation}
{d {\bf p} \over dt} = q ({\bf E}_i + {{\bf v} \over c} \times {\bf B} ),
\label{eq:motion}
\end{equation}
where ${\bf E}_i$ is the electric field induced by the rotating magnetic
field ${\bf B}$ in the pulsar magnetosphere. The symbols $q,$ ${\bf v},$
${\bf p}$ and $c$ stand for particle charge, velocity, momentum and the
speed of light, respectively. Goldreich and Julian (1969) have proposed 
that a beam of electrons (or ions) is accelerated at
the polar cap by this electric field. Sturrock (1971) suggested
that this accelerated particle beam will emit gamma
rays by the curvature radiation. These high energy photons interact
with the magnetic field and generate the electron--positron pairs.
The pairs thus created are most likely to be in  higher Landau levels,
and therefore lose their perpendicular component of the momentum
through the synchrotron radiation. The synchrotron photons with
energies above $1$~MeV can further decay into electron-positron pairs in the 
pulsar magnetic fields.

The coherent curvature radiation from these secondary plasma
particles flowing along the curved magnetic field lines is one
of the important mechanisms proposed to explain the very high
brightness temperature ($ 10^{25} - 10^{30}$~K) from pulsars 
(Pacini and Rees 1970; Sturrock 1971).
If $ {\bf J}$ is the current density due to the flow of such a
plasma then the Fourier components of coherent radiation
electric field are given by (Jackson 1962)
\begin{equation}
{\bf E}(\omega ) = -i {\omega e^{i {\omega R \over c}} \over 
\sqrt {2 \pi} R c^2} \int^{+\infty}_{-\infty} dt
\int~\hat n \times (\hat n \times {\bf J} )~ e^{i \omega 
(t - {\hat n . {\bf r} \over c})} d^3 {\bf r},
\label{eq:field}
\end{equation}
where $ R $ is the distance between the observer and the
emission point and $\hat n$ is the unit vector representing the
line-of-sight.

Consider a stationary Cartesian coordinate system--$xyz$, as shown in Fig.~1, 
centered on the neutron star where $z$-axis is aligned with the
rotation axis $\hat{\Omega}.$ Assume that $\hat n,$ 
magnetic axis $\hat m$ and $\hat{\Omega}$ lie in the $x$--$z$
plane at time $t=0.$ Consider a plasma column with elliptical cross section
flowing along the rotating magnetic field lines. Let $s_0$ be the length of plasma 
column, and $\xi_0$ and $\eta_0$ be the major and minor axes 
of its elliptical cross section, respectively. 
Let $\zeta$ be the angle between 
$\hat{\Omega}$ and $\hat n,$ and $\zeta_c$ be the angle between
$\hat{\Omega}$ and the center of momentum (CM) velocity of plasma particles $\hat v_c.$ 
The observer receives radiation when the angle between $\hat n$ and $\hat v_c$ is
$\leq 1/\gamma,$ where $\gamma$ is the Lorentz factor of
the plasma bunch. The phase $\Omega t$ represents the instantaneous position 
of the magnetic axis $\hat m.$ The instantaneous plane of the CM orbit of a
bunch is assumed to be an arc of a circle which makes an angle $\theta $ 
with the line-of-sight $\hat n.$

In the co-moving frame-$x'y'z'$, plasma waves $(\omega_p',k_p')$ 
are excited in the plasma column by the plasma instabilities such as the
oscillating-two-stream instability (Ruderman and Sutherland 1975). 
Since plasma waves are longitudinal, they are capable of creating density 
fluctuations, which can behave like particle bunches. 
Let $\rho $ be the instantaneous radius of curvature of the trajectory of
particles in the CM plane and $\mu$ be the angle along the arc of the 
trajectory measured with respect to $y'$ axis which is used in the volume integration 
appeared in Eq.~(\ref{eq:field}), and $\hat e_{\mu} $ the unit
vector tangent to the trajectory. In the co-moving frame the current 
density and the charge density are given by
\begin{equation}
{\bf J}' = \hat x' J_0' \sin (k_p'x' - \omega_p' t'), ~~~~~~~~~~~~~~~~~~
\sigma ' = \sigma_p' \sin(k_p' x'-\omega_p' t').
\label{eq:jp}
\end{equation}
Using the Lorentz transformation,  we transform the current density and 
charge density into rest framei, and substitute ${\bf J} $ into 
Eq.~(\ref{eq:field}). Next by integrating it, we obtain
\begin{equation}
{\bf E}(\omega ) = - i {\omega e^{i \omega R/c} \over \sqrt {2 \pi } R c^2} 
{\bf A},
\label{eq:electric}
\end{equation}
and
\begin{equation}
{\bf A} = {J_0 N s_0 \xi_0 \eta _0 \over 2 i }\,\, {\sin [(k-k_p)s_0 / 2] \over
(k-k_p)s_0 /2 }\,\, {\sin ({k \eta_0 \theta / 2 }) \over k \eta_0 \theta / 2}
\left \{ \hat \epsilon \theta \left ( {6 \rho ^2\over \omega c^2} \right )^{1/3}
L_1(z) - \hat y \left ( {36 \rho \over \omega ^2 c} \right)^{1/3} L_2 (z) \right \},
\label{eq:a}
\end{equation}
where $\hat \epsilon = \hat n \times \hat y',$ and $(\omega, k)$ are the
radiation frequency and wave number, and
\begin{equation}
{\omega_p' \over \omega_L} = {\kappa \gamma \over 6} + {1 \over 2 \gamma }~ ,
\label{eq:frequency}
\end{equation}
and the constant $\kappa$ is of the order $10^{-3}$ (Buschauer and Benford 1976). 
The parameter
\begin{equation}
z  =  \left ({6 \omega ^2 \rho ^2 \over c^2}\right )^{1 /3}\left
[{1 \over 2 \gamma ^2} + {\theta ^2 \over 2} - {\omega _p' \over
\gamma \omega } - {\xi_0 \over \rho } \right ],
\label{eq:z}
\end{equation}
and for positive z, we have
\begin{eqnarray}
L_1(z) & = & {2 \over 3} z^{1/ 2} K_{1 \over 3} [2 (z/3)^{3/2}], \\ 
L_2(z) & = & i {2 \over 3^{3/2}} z K_{2 \over 3}[2 (z/3)^{3/2}].
\end{eqnarray}
The functions $K_{{1 \over 3}} $ and $K_{{2 \over 3}}$ are the
modified Bessel functions.

\subsection {Calculation of $\rho $ and $\theta $ }

In the previous section we derived an expression for the Fourier 
components of the
radiation electric field ${\bf E}(\omega)$ as a function of the
radius of curvature $\rho $ and the angle $\theta.$ Now, we 
shall introduce a method to estimate these quantities in the
$xyz$ frame as functions of observable quantities such as  
$\zeta,$ $\Omega t$ and the magnetic axis
inclination angle $\alpha.$ The 
angular width of open field line region above the polar cap varies as
$ w = (r \Omega / c)^{1/2} = ({r / R_{LC}})^{1/ 2},$ where $R_{LC}$
and r are the light cylinder radius and emission height, respectively. 
These are the field lines, from which the coherent radio waves
are expected to be produced.  Blaskiewicz et~al. (1991) have showed
that, in the first order of calculations, the terms of the order
of $w^2$ are negligible in the equation of motion
(Eq.~\ref{eq:motion}).

We relax the assumption of confining the emission region to a
constant radius made by Blaskiewicz et~al. (1991), and calculate
velocity and acceleration in the CM frame of
particles along the portion of magnetic field lines from which the radiation 
is receivable. At an arbitrary time $t,$ the magnetic 
axis can be represented as
\begin{equation}
\hat m = \sin\alpha[\hat x\cos(\Omega t)+\hat y\sin(\Omega t)]+
\hat z\cos \alpha .
\label{eq:magnet}
\end{equation} 
In a co-rotating magnetosphere, the guiding center velocity of the 
center of momentum is
\begin{equation}
{\bf v}_c \cong v_{\|} \hat b + {\bf \Omega} \times {\bf r} , 
\label{eq:veloc}
\end{equation}
where $v_{\|}$ is the velocity of the center of momentum
parallel to $\hat b,$ the unit vector tangent to the dipolar magnetic
field lines (Hibschman and Arons 2001).

The condition for receiving the radiation is that line-of-sight
should lie inside the particle radiation beam with angular width
$2/\gamma,$ which is met when $\zeta _c \simeq \zeta.$
This helps us to write $\hat v_c $ as
\begin{equation}
\hat v_c = \sin \zeta \,\cos \phi \,\hat x \,+\, \sin \zeta \,\sin \phi \,\hat y +
 \cos \zeta \,\hat z
\label{eq:vc}
\end{equation}
Since $| \hat v_c - \hat m |$, $| \hat v_c - \hat r |$ and $|
\hat m - \hat r |$ lie inside the open field line region and all are, 
of the order of $w$. Therefore, we have
\begin{equation}
\hat r = {2 \over3 } \left ( \hat v_c - {r \over c } {\bf \Omega} \times \hat v_c \right )
\left (1 + {\epsilon ^2 \over 18}\right ) + {1 \over 3} 
\left (1 + {2 \epsilon^2 \over 9 }\right ) \hat m +O(w^3),
\end{equation}
where
\begin{equation}
\epsilon = \left [ \sigma ^2 + 2 \sin \alpha \, \sin \zeta \, \{1-\cos(\Omega\, t - \phi)\}\right] ^{1/2}~,
\label{eq:epsilon}
\end{equation}
$\sigma = \zeta  - \alpha $ is the impact parameter
and $\phi$ is the angle between $\bold {v_c}$ and $\hat n.$
Substituting this result for $ \hat r $ into
Eq.~(\ref{eq:veloc}) gives the center of momentum velocity
${\bf v}_c.$ Differentiating ${\bf v}_c $ with respect to time
gives the acceleration 
\begin{equation}
{\bf a} = -{1 \over 2r } [\hat m \,c^2 - \hat v_c (\hat v_c . \hat m)\, c^2 - 3\, c\, r\, \Omega (\hat z \times \hat v_c ) ] + O({w^3}).
\label{eq:acceleration}
\end{equation}
The instantaneous radius of curvature of a particle orbit is given
by
\begin{equation}
\rho = {v^2_c \over a } \cong {2 r \over [\epsilon ^2 - \left (6 r \Omega / c\right )
\sin \alpha \,\sin \zeta \,\sin(\Omega \,t - \phi )]^{1/2} }.
\label{eq:rho}
\end{equation}
The angle $\theta $ between  the line-of-sight and the instantaneous
plane of orbit is given by
\begin{equation}
\sin \theta  = \hat n \,.\, (\hat v_c \times \hat a )
\label{eq:theta}
\end{equation}
or
\begin{eqnarray}
|{\bf a}| \, \sin \theta & = & \sin\zeta \,[\sin\zeta\, \sin\phi
\, \cos\alpha - \cos\zeta \, \sin\alpha
\sin(\Omega t)] \nonumber \\
& + & \sin\zeta \, \cos\zeta \, [\sin\alpha \, \cos\phi
\,\sin(\Omega t) - \sin\alpha \,\sin\phi \,\cos(\Omega t)]
\nonumber \\ & -& 3 r \Omega \cos \beta \sin ^2 \beta [1 - \cos
(\Omega t) ] / c \, ,
\end{eqnarray}
where $|{\bf a}|$ is the magnitude of acceleration. Equations (16)
and (18) specify $ \rho $ and $\theta $ as functions of pulse
phase.  Substituting these relations into Eq.~(\ref{eq:field})
gives ${\bf E}(\omega )$ as a function of the rotation phase.
From Eqs.~(\ref{eq:rho}) and (18) we infer that:

a) If $\rho_l$ is the curvature of particle trajectory on the leading
side ($\Omega t - \phi < 0 $) and $\rho_t$ is the curvature on the trailing
side ($\Omega t - \phi > 0 $) then Eq.~(\ref{eq:rho}) shows that $\rho_l<\rho_t.$

b) For phases $\pm \phi $ on either sides of the orbital
plane, which is at $\phi = 0,$ Eq.~(18) shows that $\theta $ is asymmetric, i.e.,
the values of $\theta $ are not same at $\pm \phi .$

\subsection{Stokes parameters}

Let $E_y(\omega) $ and $E_{\epsilon }(\omega)$ be the components
of radiation electric field ${\bf E} (\omega )$ given by Eq.~(4).
Then the Stokes parameters can be defined as
\begin{eqnarray}
I & = & E_y\, E_y^* + E_{\epsilon} \,E_{\epsilon }^* \nonumber
\\ & = & W^2 D^2 S^2 \left \{{4 \over 27}
\left ({36 \rho \over \omega^2c}\right)^{2/3}\!\! z^2K_{2 \over 3} 
[2 (z/3)^{3/2}]^2+\theta^2\left({6\rho^2\over\omega c^2}
\right)^{2/3}\!\! \left({4 \over 9}\right) z K_{1 \over 3} [2 (z/3)^{3/2}]^2 
\right \},~~~ \\ \nonumber 
\\ Q & = & E_y\, E_y^* - E_{\epsilon}\,
E_{\epsilon }^* \nonumber \\ & = & W^2 D^2 S^2 \left \{{4
\over 27}
\left ({36 \rho \over \omega ^2c }\right )^{2/3}\!\! z^2 K_{2 \over 3} [2(z/3)^{3/2}]^2
-\theta ^2 \left ({6 \rho ^2\over \omega c^2 }\right )^{2/3}\!\! \left({4
\over 9}\right) z K_{1 \over 3} [2 (z/3)^{3/2}]^2 \right \},  \\ \nonumber 
\\ U & = & 2\, {\rm Re} (E_y\, E_{\epsilon}^*) = 0, \\ \nonumber 
\\ V & = & 2\, {\rm Im} (E_y \,E_{\epsilon}^*) \nonumber \\ &
   = & - 2 W^2  D^2 S^2 \left({36 \rho \over \omega ^2c
}\right)^{1/3}\left ({6 \rho ^2\over \omega c^2 }\right)^{1/3}
\left ({4 \over 3^{5/2}} \right ) \theta z^{3/2} K_{2 \over 3} [2(z/3)^{3/2}]
K_{1 \over 3} [2 (z/3)^{3/2}],
\end{eqnarray}
where 
\begin{equation}
W = {J_0 N_0 S_0 \xi_0 \eta_0 \over 2},\,\,\,\,\,  D={\sin({k \eta_0 \theta / 2 }) \over {k \eta_0 \theta / 2 }}
,\,\,\,\,\,   S={\sin[{(k-k_p) s_0 / 2 }] \over {(k-k_p) s_0 / 2
} } ,
\end{equation}
where $N_0$ is the number of radiating bunches in a column. 
 The polarization position angle is defined as
\begin{equation}
\psi = {1 \over 2} \tan^{-1}\left ({U \over Q}\right).
\end{equation}
Since U is zero, $ \psi $  can be zero or $ \pi / 2. $
However, $\psi = \pi /2$ represents the real orientation of
${\bf E}(\omega)$, as predicted by Eq.~(2).  Note that 
due to the special choice of the coordinate system-$\hat \epsilon\hat y\hat n$
attached to the center of momentum, $\psi $ becomes constant for a given 
field line. But when the line-of-sight moves from field line to field line 
due to rotation, polarization position angle swings in agreement with 
the rotation vector model (Radhakrishnan and Cook 1969). In the $xyz$ coordinate
system, the position angle is the angle between the radiation
electric field and the projected spin axis on the plane
of sky. Since the particle acceleration ${\bf a}$ is parallel to ${\bf E},$ we shall use
${\bf a}$ as the reference for convenience.  Since $\hat n$ is normal to
both the plane of sky and $\hat y,$ the plane of sky contains the $y$--axis. 
Hence the linear polarization angle $\psi$ can be estimated from
\begin{equation}
\tan \psi = {\hat a\,.\,\hat {y} \over \hat a\,.\,(\hat n \times \hat y) }.
\end{equation}
After substituting for $\hat a,$ it can be simplified as
\begin{equation}
\psi = \tan ^{-1} \left[ { 3\, (r \Omega/c) \, \sin \zeta - \sin\alpha
 \sin (\Omega t) \over \sin \sigma  + \sin \alpha  \cos \zeta
[1 - \cos (\Omega t)] } \right].
\label{eq:psi}
\end{equation}

\section{ Numerical calculation}

For numerical calculations we adopt the values of emission
heights and component locations provided recently by Gangadhara
and Gupta (2001) for PSR B0329+54 at 606 MHz. Even though they have
proposed nine emission components, we consider only three strong
components marked as i, iii  and iv in Fig.~4a, which have
higher polarization. Out of these three, iii is a core
component, and i and iv are the conal components of cone
number 3 (Gangadhara and Gupta 2001). The phase locations of the
components i and iv are $-12.6 \degr  \pm 0.64 \degr $ and
$9.5 \degr \pm 0.64 \degr $, respectively. They assumed zero
pulse phase for the core, however, the core is expected to be produced at
some height above the polar cap and most likely  shifted from
zero pulse phase to the trailing side due to aberration and retardation. In
our calculation we assume core component is emitted at an height of $ r_{
\rm core} \simeq 2 r_{\rm NS} \simeq  20 $ km above
the polar cap.

The dipole magnetic field strength of PSR B0329+54 is about $B_0
= 1.2 \times 10^{12}$ G at the surface of the neutron star.
Using the characteristic curvature radiation frequency given by
$\omega _c = (3/2) \gamma ^3 (c/\rho), $ 
we can estimate the particle Lorentz factor, $\gamma$. For $\nu _c =
\omega _c/(2 \pi) = 606 $ MHz, emission height $r_{\rm cone} = 600 \pm 180$
km, Eq.~(16) gives $\rho _l \simeq 10^4 \pm 10^3$ km and 
$\rho _t \simeq 2 \times 10^4 \pm 10^3 $ km.
For these values, we find $ \gamma = 300 \pm 190 $ for leading component i 
and  $ 480 \pm 60$ for trailing component iv. Having the estimates of
particles Lorentz factor, we compute the Stokes parameters using
Eqs.~(19)--(22).

Coherency factor $S^2$ (Eq.~23) becomes maximum for $ k \simeq k_p$, i.e., the
coherency is more effective when the radiation wave number is of
the order of the plasma wave number. We use this resonant matching condition in
our numerical calculation of polarization parameters.
The diffraction term $D^2 $  takes a maximum
value $\simeq 1 $ for the parameters used in our calculations.
Further, since radius of curvature is much greater than the plasma columns 
height  $\xi_0$,
the last term on the right hand side of Eq.~(7) is negligible
compared to the other terms.

Figure 2 shows the results of
the numerical estimation of intensity I, linear polarization L
$= \sqrt {Q^2 + U^2}$, circular polarization V and linear
polarization angle $\psi$. To calculate  the polarization of each 
component in this figure, we considered a bunch of field lines with a 
proper rotation phase. The main features of Fig.~2 are:

a) The asymmetry in the phase location of conal components i 
and iv with respect to the core iii arises due to the fact
that the emission height of conal components is higher than the
core component, which is explained as aberration and retardation
effect by Gangadhara and Gupta (2001).

Since Blaskiewicz et al. (1991) considered the constant emission height
they could estimate only the influence of aberration on the pulse width but not
retardation. But we have relaxed the assumption of
constant emission height by estimating the emission over the range
of height, where the conditions are conducive for the coherent curvature
emission. So, we are able to estimate the change in components width
due to both the aberration and the retardation, which is almost double
the value predicted by Blaskiewicz et al. (1991).

b) Figure 2a shows that the phase width of the component i
is broader than that of the component iv.  This broadening is
a consequence $\rho _l$ becoming smaller than $\rho
_t $. We propose that this result can have observational
evidences. In the next section we analyze the components
width of 24 pulsars to check this prediction.

c) As we expected from our discussion on radius of curvature
with similar particle densities on the field lines the
Stokes parameter, I, for trailing component becomes stronger than that of the
leading in agreement with Jackson(1962, Eq.~14.93).

d) We define the clockwise rotation of ${\bf E} (\omega)$ as negative circular
polarization (V$ < 0$) and counter clockwise rotation as positive
circular polarization (V$> 0$) then Fig.~2a shows that all the three
components (i, iii and iv) display  the well known 
anti--symmetric type of circular polarization.  However, there is a
difference in the relative magnitudes of circular polarization
on either sides of each component center, i.e., positive V
dominates over negative V in the case of component i while
in component iv negative V dominates. This happens because
of the fact that, due to the inclination of plane of field lines
with respect to rotation axis, while observing say component
i we tend to receive more radiation from the leading side of
the plane of any given field line (associated with that
component) compared to that received from the trailing side
of the same field line.

In the case of core component the planes of magnetic field lines
are nearly parallel to the rotation axis. Therefore we tend to
receive equally both positive and negative parts of circular
polarization from either sides of field line planes. Hence
circular polarization becomes anti--symmetric for the core component.

Further the circular polarization is affected by rotation in
such a way that in the case of leading component it makes
positive V weaker but in the case of trailing component it enhances
negative V. The simulated curve representing V in Fig.~2a depicts this behavior.

e) Figure 2b shows the polarization angle swing with respect to
pulse phase. It shows that the centroid of polarization angle
curve shifted toward the trailing side. This shift arises due to the rotation
as proposed by Blaskiewicz et~al. (1991). However, Hibschman and Arons (2001)
have shown that current flow above the polar cap can shift polarization sweep 
in the opposite direction. 

f) For the purpose of comparison we repeated the calculation of Stokes
parameters by changing the impact parameter to $-2.5 \degr $, and plotted in 
Fig.~3b.
We notice that the change of
sense of impact parameter $\sigma $, flips the
polarization angle swing and sense of circular polarization.

\section{Observational evidences }

Our model predicts that due to rotation
the width of leading component becomes broader than the width of
the trailing component. Further, the circular polarization on the leading
component dominates with one sense while on the trailing component the 
other sense dominates. To find the observational
evidences in this regard we analyzed the average profiles of about 24 pulsars
and also the single pulse polarimetric data of PSR~B0329+54.
\begin{deluxetable}{lclll}
\tablecaption{Pulse width comparison \label{tbl-1}}
\tablewidth{0pt}
\tablehead{
\colhead{PSR~B} & \colhead{Frequency (MHz)}   & \colhead{Period (s)} &
\colhead{FWHM$_l$/FWHM$_t$} &
\colhead{Reference\tablenotemark{*}}
}
\startdata
 0052$+$51   &    610         &  2.115 & 1.032$\pm$2.20E$-$02   &  gl98      \\ 
 0059+65     &    610         &  1.679 & 1.052$\pm$2.22E$-$02   &  gl98      \\
 0148$-$06   &    610         &  1.465 & 1.136$\pm$7.49E$-$03   & gl98       \\ 
 0226$+$70   &    610         &  1.467 & 1.3574$\pm$0.0         &  gl98      \\ 
 0301$+$19   &    610         &  1.388 & 0.937$\pm$0.0          &  gl98      \\
 0525$+$21   &    610         &  3.746 & 1.056$\pm$0.0          &  gl98      \\ 
 0559$-$05   &    610         &  0.396 & 1.772$\pm$0.0          &  gl98      \\ 
 0818$-$41   &    660         &  0.545 & 1.319$\pm$2.85E$-$02   &  qmlg95    \\
 0834$+$06   &    4800        &  1.274 & 0.868$\pm$2.0          &  kkwj97    \\ 
 1254$-$10   &    610         &  0.617 & 1.106$\pm$2.59E$-$02   &  gl98      \\ 
 1601$-$52   &    138         &  0.658 & 1.178$\pm$6.24         &  qmlg95    \\
 1648$-$17   &    606         &  0.973 & 1.190$\pm$0.0          &  gl98      \\ 
 1800$-$21   &    1642        &  0.134 & 0.850$\pm$3.58E$-$02   &  gl98      \\ 
 1822$-$09   &    610         &  0.769 & 1.141$\pm$0.0          &  gl98      \\
 1823$-$13   &    1642        &  0.101 & 1.190$\pm$9.98E$-$02   &  gl98      \\ 
 1839$-$04   &    606         &  1.840 & 1.442$\pm$9.10E$-$03   &  gl98      \\ 
 1914$+$09   &    610         &  0.270 & 1.628$\pm$0.0          &  gl98      \\
 1919$+$21   &    610         &  1.337 & 1.111$\pm$0.0          & gl98       \\ 
 1935$+$25   &    606         &  0.201 & 0.839$\pm$0.0          &  gl98      \\ 
 1942$-$00   &    646         &  1.046 & 1.354$\pm$3.90         &  kmj96     \\ 
 2020$+$28   &    610         &  0.343 & 1.150$\pm$0.0          & gl98       \\ 
 2044$+$15   &    610         &  1.138 & 1.572$\pm$1.45E$-$02   &  gl98      \\ 
 2053$+$21   &    606         &  0.815 & 0.920$\pm$0.0          &  gl98      \\ 
 2224$+$65   &    610         &  0.682 & 1.025$\pm$8.98E$-$03   & gl98       \\  
\enddata
\tablenotetext{*}{gl98: Gould and Lyne (1998);~~kmj96: Kaspi et~al. (1996);
~~kkwj97: Kijak et~al. (1997);~~qmlg95: Qiao et~al. (1995)} 

\end{deluxetable}

\subsection{Effect of the rotation on component width }
We estimated the components width on trailing and leading sides of
the average profiles of 24 pulsars using the data available in EPN
format on web.  We compared the full width at half maxima for leading
component (FWHM$_l$) and trailing component (FWHM$_t$) of each pulsar. To select our
sample we considered the following criteria:  a) The pulse
profile should not change drastically over the range of 600
MHz to 1000 MHz, i.e., the pulse
profile should be stable through out this range of frequencies.
b) The conal components should be prominent and not affected by the
presence of core component.  c) The leading and trailing
components should be distinguishable enough, so that we can fit
Gaussians to them without ambiguity.

According to Rankin's (1990) classification the best candidates
for our sample are conal Double, Triplet and Multiple pulsars.
We choose 8 of the pulsars referred as Double conal in Rankin
(1990), and not considered the sources like PSR~B1133+16 because
they may have more weaker components than the visible components in
the average pulse profiles (Nowakowski 1996).
We found the  FWHM$_l$ and FWHM$_t$ by fitting Gaussians to the pulse
components, and estimated their ratios. We did this analysis at
different frequencies for each
pulsar and presented only the results at 610 MHz, except for a few
sources for which the data was not available at this frequency.
Table.~1 shows the sample and ratio of the FWHM of leading to
trailing components for a set of 24 pulsars.  It shows that 19 out 
of 24 pulsars have broader leading
components than trailing one.  Of course, for sources like 
PSR~B0834+06 most likely a core component exists close to the
trailing component and consequently it is broader than the leading
component.
 
We performed Kolmogorov--Smirnov test on the two data sets
comprising of one for the leading side components width, and the other for
trailing side components width. The mean component widths were 5.83$\degr$ and 5.06$\degr$
for the leading and the trailing components, respectively.
The Kolmogorov-Smirnov statistic D was found to be 0.17$\degr ,$
and the significance level P was $0.86\degr .$ The 95\% confidence interval for
the actual mean widths were 3.24$\degr$ and 2.71$\degr$ for the
leading and trailing components, respectively.

\subsection{Interpretation of circular polarization from PSR B0329+54}

To study the nature of circular polarization from PSR B0329+54,
we analyzed the single pulse data at 606~MHz taken on August 30,
1996 from the Lovell telescope at Jodrell Bank. We considered
about 2500 single pulses with time resolution of 0.249 ms.  The
average polarization parameters: intensity~I, linear~L and
circular~V polarization are plotted as functions of pulse
phase in Fig.~4a. The continuous line curve indicates I while
the broken and dotted ones represent L and V, respectively,
in arbitrary units.  The average polarization angle (bullet
symbols) is given in Fig.~4b.

The gray-scale maps in Fig.~4b-d show the frequency of
occurrence of polarization parameters with respect to the pulse
phase in single pulses. We used the {\it pgplot} routines developed by Pearson
(1989) for making gray-scale maps.  The Fortran subroutine {\it
pggray} draws gray-scale map of an array of the polarization
parameters vs phase, by determining the shade of each
point from the corresponding array value. The shade is a number
in the range from 0 to 1 obtained by {\it linear interpolation}
between the background level (white) and the foreground level
(black). The white regions in the maps are with shade~=~0 and
darkest regions are with shade~=~1. This technique has become a
powerful tool in analyzing the pulsar polarization properties
(e.g. Stinebring et~al. 1984a,b). The panels (b), (c) and (d) in
Fig.~4 represent the polarization angle, linear and circular
polarization gray-scale maps, respectively.

The darkest shades represent the most probable regions of
occurrence.  The gray-scale maps were made from all those phase
bins where the linear polarization L is above $4\sigma$ level.
Here $\sigma$ is the {\em rms} of L in the off pulse region.
All those phase bins, where the condition 
${\rm L}^2+{\rm V}^2\leq {\rm I}^2$ was not met, were excluded 
as they lead to spurious polarization quantities.

The average circular polarization was obtained from the
gray-scale map (Fig.~4d) and the continues curve superposed in
Fig.~4d. Note that for
the sake of plotting on the selected scale, the whole curve has been
amplified by a factor of 5.  It shows under the leading component
i it is positive and negative under the trailing component
iv while it is anti-symmetric under the core.  This type
circular polarization can be explained by considering the
dipolar magnetic field lines with inclined magnetic axis with respect 
to the rotation axis. In agreement with this observational result, 
our Fig.~2a shows the dominance of positive
circular polarization under the leading component and negative
dominance under the trailing component, and anti-symmetric type
under the core. 

\section {Conclusion}

We have calculated the Stokes parameters I, Q, V for pulsars radio
emission by taking into account the effect of the neutron star's
spin for the first time. Figures 2a and 3a shows that under the influence of 
rotation, intensity I for the leading component becomes less than 
that of the trailing component for a similar  distribution of 
emitting particles on different field lines.

According to the conal model (Rankin 1983a,b, 1990, 1993) the
emission regions on the polar cap are organized in concentric hollow
cones. Our simulation shows that the leading component becomes
broader than its trailing counterpart.  This broadening is induced by rotation
through change in curvature of particle trajectories, as discussed 
in section 3. By analyzing 24 pulsar pulse profiles we find that 19 of them
have leading components broader compared to trailing ones, and thereby 
confirm the possibility of detecting such an effect through observations. 
We note that such a broadening becomes observable only when the emission 
components are organized in the form of nested cones. 

We have found, due to the inclination of magnetic axis with respect to 
the rotation axis and alteration of the particles trajectory by rotation, one
sense of circular polarization becomes stronger in conal components of 
single pulses. Due to this enhancement of one sense of circular polarization
in single pulses, one sense of circular polarization survives in 
the conal components of average pulse profile.
It is worth mentioning that the inclination of magnetic field planes 
with respect to the line-of-sight is mainly responsible for the enhancement 
of one sense of circular polarization in conal component. But in the 
case of core component this effect vanishes and consequently lead to the 
circular polarization with an anti-symmetric profile.

Radhakrishnan and Rankin (1990) found a strong correlation between the senses 
of polarization angle swing and circular polarization.  Our
simulation shows that changing the sign of impact parameter flips the sense 
of circular polarization and polarization angle swing. Therefore, the above  
correlation is caused by change in the sign of the impact parameter.  
The centroid of polarization angle curve does not coincide with the 
pulse center as a consequence of rotation.

\acknowledgments

We thank Y. Gupta and V. Krishan for carefully reading the manuscript and making 
comments. Also, we are grateful to A. G. Lyne for providing the Jodrell Bank data 
of PSR B0329+54. Part of this research has made use of the EPN data
base maintained by the Max--Planck Institut f\"ur Radio Astronomie on web.

\clearpage

\begin{figure}
\plotone{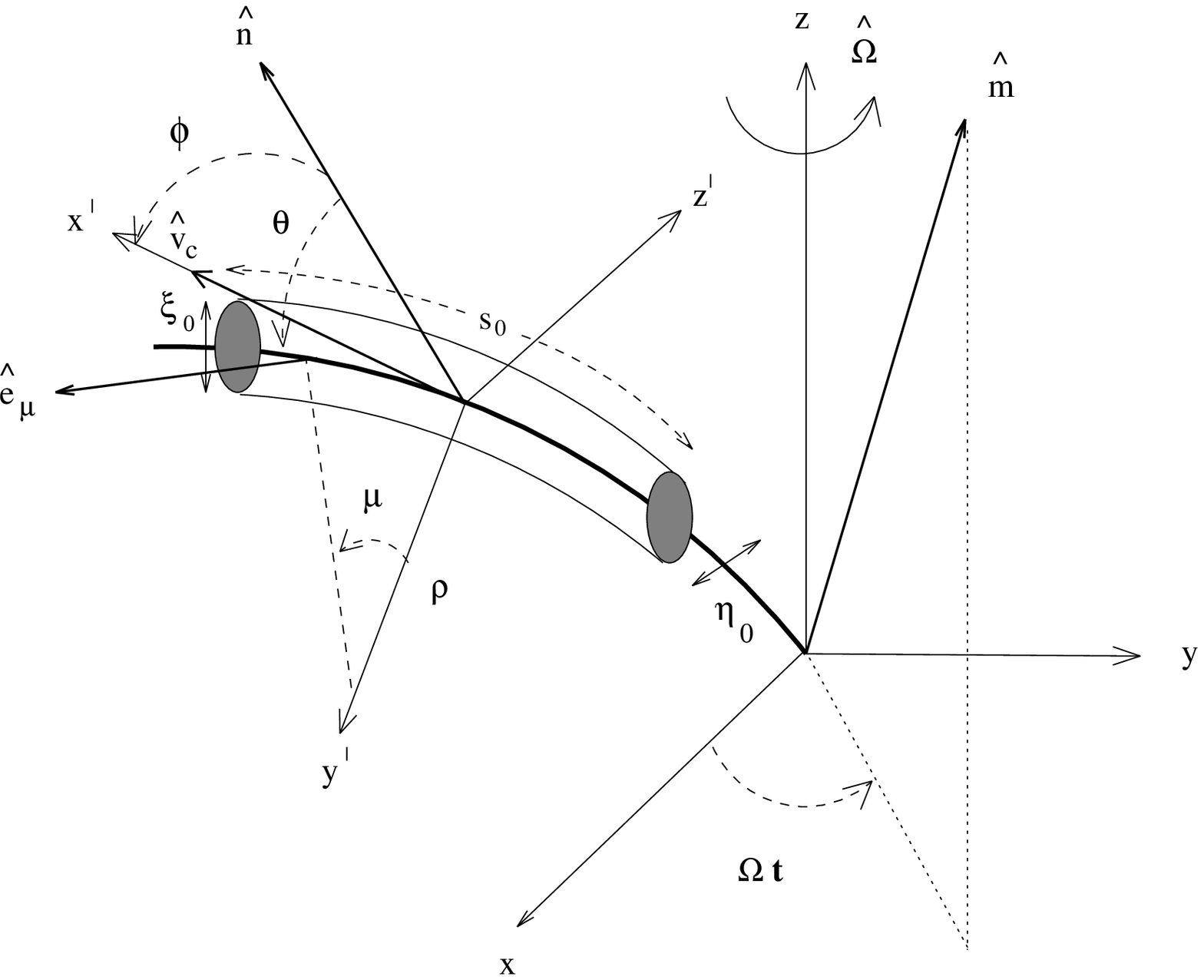}
\caption{Emission region geometry. The lab frame--$xyz$ 
centered on the neutron star and $x'y'z'$ is the co-moving reference frame. 
The column of plasma with length $s_0$ and elliptical cross section with major 
axis $\xi_0$ and minor axis 
$\eta_0$ contains $N_0$ bunches of coherently radiating particles.  
$\hat n$ and $\hat m $ are the line of sight and magnetic dipole unit vectors.
 \label{fig1}}
\end{figure}

\begin{figure}
\plotone{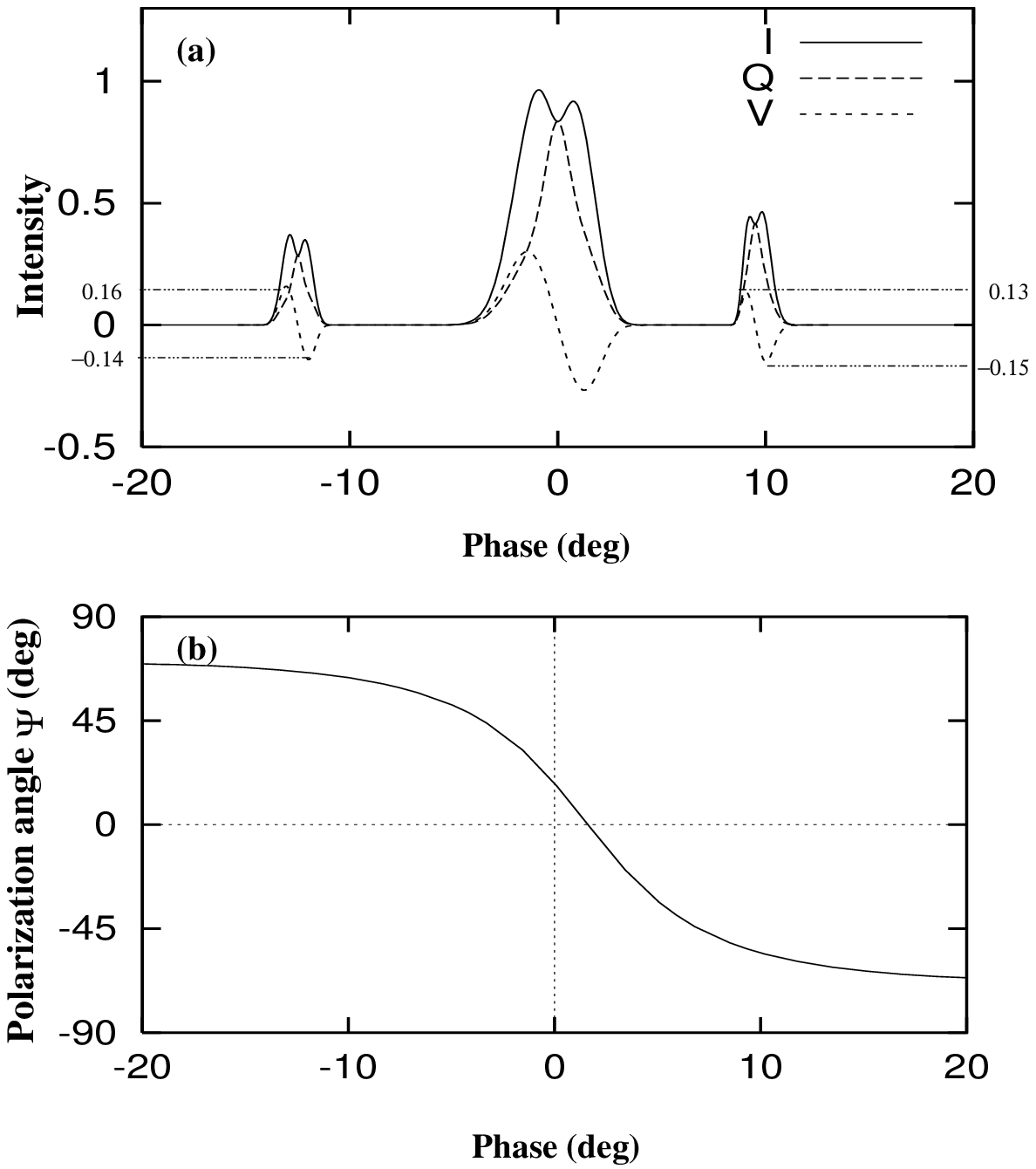}
\caption{ (a) Normalized Stokes parameters, I, L, V, and (b) polarization angle 
for a single pulse simulated using $\alpha = 30 \degr$, $\sigma = 2.5 \degr $, 
$\omega_L = \omega _r /3$
and $\omega _r = 610$ MHz.\label{fig2}}
\end{figure}

\begin{figure}
\plotone{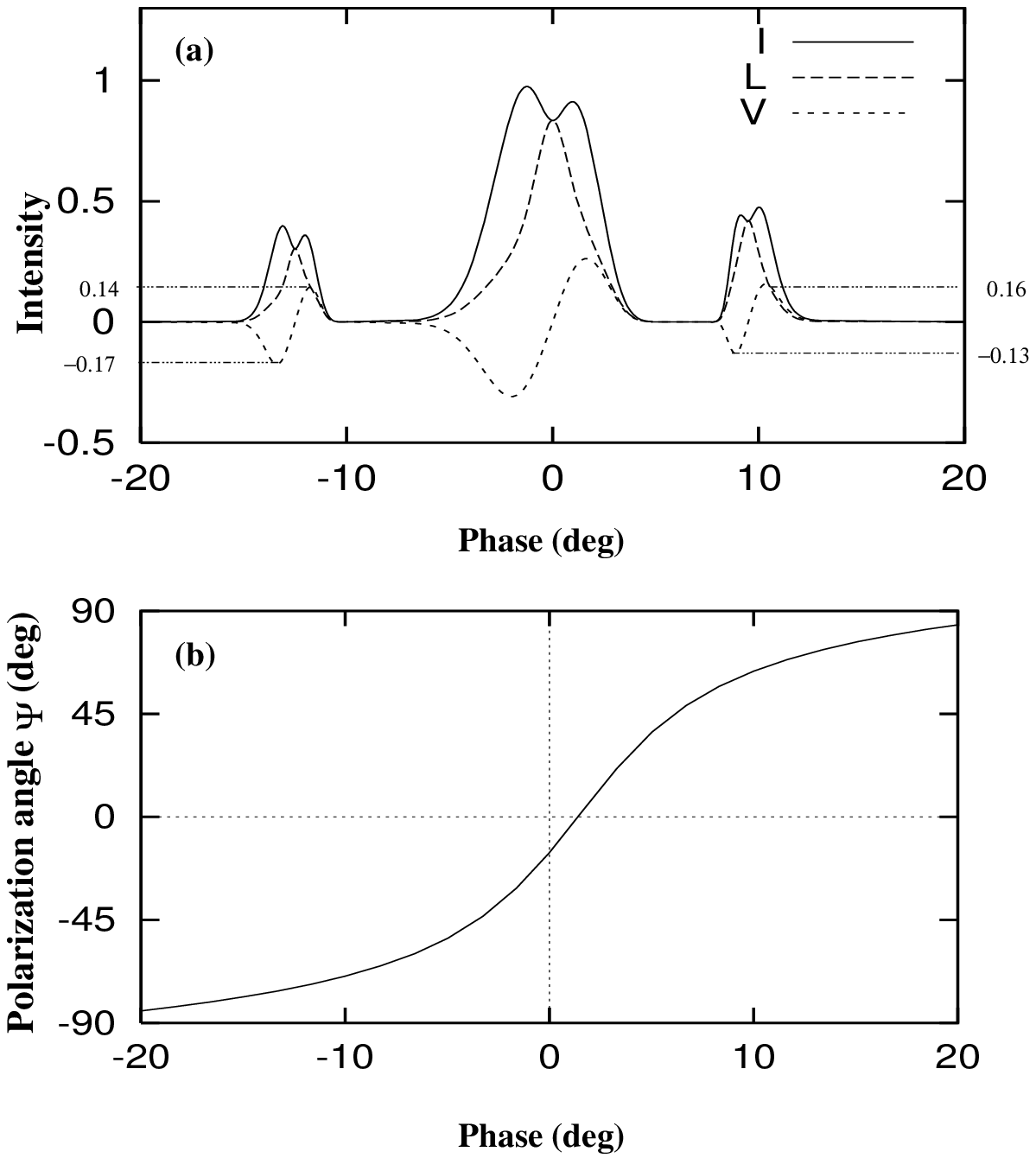}
\caption{ (a) Normalized Stokes parameters, I, L, V, and (b) polarization angle       
for a single pulse simulated using $\alpha = 30 \degr$, $\sigma = - 2.5 \degr $, 
$\omega_L = \omega _r /3$
and $\omega _r = 610$ MHz.\label{fig3}}
\end{figure}

\begin{figure}
\plotone{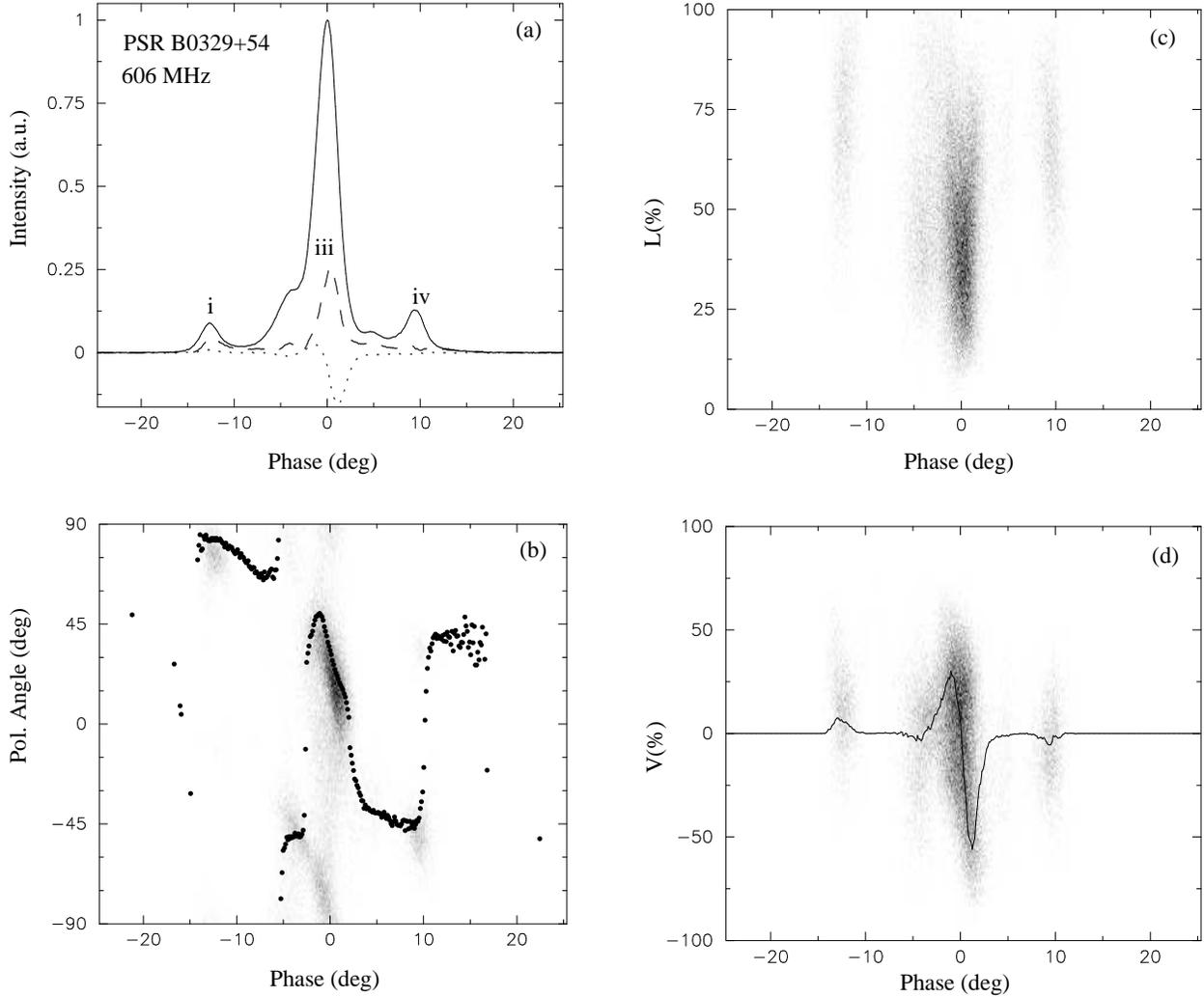}
\caption{ (a) The average pulse profile of PSR~B0329+54 with arbitrary 
intensity units (a.u), and (b) average polarization angle
$(\psi)$ (bullet symbols), and polarization angle
gray-scale map obtained from individual pulses. The panels (c)
and (d) represent the gray-scale maps of $L(\%)$ and $V(\%).$
The shade is a number in the range 0 (white) and 1 (black)
obtained by linear interpolation between the background and
foreground levels. The average circular polarization curve
(solid line) is superposed in the panel (d).  \label{fig4}}
\end{figure}

\end{document}